\documentclass[conference,10pt,a4paper]{IEEEtran}
\usepackage[T1]{fontenc}
\usepackage[a4paper,total={184mm,239mm}]{geometry}
\usepackage{cite}
\usepackage{amsmath,amssymb,amsfonts}
\usepackage{graphicx}
\usepackage{textcomp}
\usepackage{xcolor}
\usepackage{booktabs,array,tabularx}
\usepackage[hidelinks]{hyperref}
\hypersetup{pdfauthor={Ruiyang Ma, Sheng Qi, DaiKang Kuang, Teng Ma, Guojie Luo}}
\definecolor{inkblue}{RGB}{35,75,106}
\definecolor{lightblue}{RGB}{233,242,248}
\definecolor{lightgreen}{RGB}{236,245,237}
\definecolor{mutedred}{RGB}{157,52,40}

\begin{document}
\title{Compositional Performance Modeling of LLM Inference Clusters with Hierarchical KV Caching}
\author{%
\IEEEauthorblockN{Ruiyang Ma\IEEEauthorrefmark{1},
Sheng Qi\IEEEauthorrefmark{1},
DaiKang Kuang\IEEEauthorrefmark{1},
Teng Ma\IEEEauthorrefmark{2},
Guojie Luo\IEEEauthorrefmark{1}}
\IEEEauthorblockA{\IEEEauthorrefmark{1}\textit{Peking University}\qquad
\IEEEauthorrefmark{2}\textit{Independent Researcher}\\
ruiyang@stu.pku.edu.cn\quad shengqi2018@pku.edu.cn\quad kuangdaikang@stu.pku.edu.cn\\
stmatengss@gmail.com\quad gluo@pku.edu.cn}
}
\maketitle
\thispagestyle{empty}
\begin{abstract}
Design-space exploration of LLM inference clusters requires predicting how hardware and serving policies interact with hierarchical key--value (KV) caching. Isolated operation profiles alone do not determine cluster latency: reuse changes computational demand, while KV location, capacity, and contention govern execution order. We present a compositional discrete-event simulator based on separating local service costs from shared resource and KV availability. Compute and transfer models calibrated on one-prefill/one-decode (1P--1D) deployments are reused across compatible workers, enabling policy and topology comparisons without per-configuration latency fitting. Validation against SGLang covers Qwen3-8B and Qwen3-32B on 1P--1D and 3P--1D clusters. Mean request end-to-end latency error is 1.34\% for question answering and 6.36\% for captured coding sessions, measured as mean absolute percentage error. CPU simulation reduces replay turnaround by 15.33$\times$ on average; real replay includes prescribed arrival gaps. Measured session-affinity and simulated retention studies demonstrate how the framework guides policy selection by predicting the joint effects on computation, latency, and KV occupancy.
\end{abstract}
\section{Introduction}
Designing an LLM inference cluster requires choosing hardware resources and serving policies together. Prefill--decode disaggregation provisions the two phases independently, while hierarchical key--value (KV) caching and locality-aware routing reduce repeated computation~\cite{distserve,mooncake}. Evaluating these choices on physical deployments is costly, motivating simulation for early design-space exploration.

Existing inference simulators predict operation costs from hardware profiles and compose them with serving models~\cite{vidur,llmservingsim}. With hierarchical caching, however, operation costs alone leave a crucial question unresolved: when can a request use the KV it needs? Routing and retention change both computational demand and data availability. Evaluating their interaction requires composing service costs with cache state and resource dependencies.

The challenge is that a cached prefix is not necessarily ready for computation. Routing determines which worker can access it, and eviction determines whether it remains available~\cite{sglang,mooncake}. A host-memory copy requires device capacity and retrieval before use. Concurrent requests further alter queues and batches. The simulator must therefore track reusable content, physical capacity, and execution readiness together.

Our insight is to separate local service costs from the shared state that determines execution eligibility. We present a compositional simulator that distinguishes logical KV content, reserved capacity, and replicas ready for computation. An event-driven orchestrator couples calibrated compute and transfer models through this state and resource queues. Models calibrated on a logical one-prefill/one-decode (1P--1D) deployment can then be reused across compatible workers, enabling policy and topology changes without per-configuration end-to-end latency fitting.

\begin{figure}[!t]
\centering
\includegraphics[width=0.96\columnwidth]{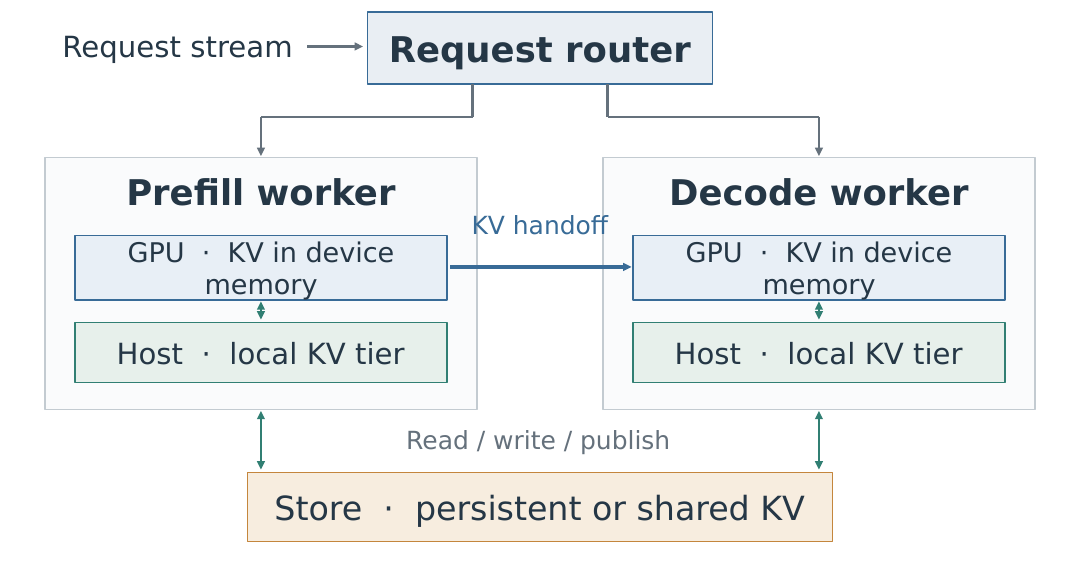}
\caption{KV hierarchy: device memory serves computation; host and optional storage tiers extend reuse through retrieval.}
\label{fig:cluster}
\end{figure}

Validation against SGLang~\cite{sglang} covers Qwen3-8B and Qwen3-32B~\cite{qwen3} in 1P--1D and 3P--1D deployments. Mean absolute percentage error (MAPE) in request end-to-end (E2E) latency averages 1.34\% for question answering (QA) and 6.36\% for captured coding sessions, with replay cases weighted equally. CPU simulation reduces replay turnaround by 15.33$\times$ on average; real replay includes prescribed arrival gaps. Session-affinity and time-to-live (TTL) studies demonstrate policy evaluation using the same service models.

The paper makes three contributions:
\begin{itemize}
\item We formulate KV-aware service composition through explicit residency, capacity reservations, and completion events.
\item We develop analytical compute and transfer models that reuse local calibration across compatible cluster workers.
\item We demonstrate accurate latency and reuse prediction across two topologies, with lower E2E and first-token errors than an adapted LLMServingSim on the evaluated workloads.
\end{itemize}

\section{Background and Related Work}
\subsection{Inference with Hierarchical KV Caching}
LLM inference comprises prefill, which processes uncached input tokens and produces KV state, and decode, which extends each sequence token by token. Disaggregation assigns these phases to separate workers, each using one or more accelerators. It allows independent provisioning but requires KV handoff and sufficient destination capacity~\cite{distserve}. Within a worker, continuous batching updates the active request set between iterations~\cite{orca}, while chunked prefill divides prompt processing into smaller steps~\cite{sarathi}. Paged allocation and prefix caching further shape execution by managing physical KV capacity and reusing content across requests~\cite{vllm,sglang}.

Hierarchical caching extends reuse beyond device memory by retaining KV in host memory or storage, at the cost of retrieval~\cite{flexgen,hicache}. Figure~\ref{fig:cluster} illustrates this organization. Device memory (L1) supplies computation directly; local host memory (L2) retains prefixes beyond device capacity; and optional shared or persistent storage (L3) extends retention or access across workers~\cite{mooncake}. We distinguish tiers by accessibility, so shared DRAM may serve as L3. This distinction separates the existence of a reusable prefix from the cost and feasibility of accessing it.

\subsection{Related Inference Cluster Simulators}
\textit{Execution-cost modeling.} Vidur uses operator profiles~\cite{vidur}, with prefix caching and routing policies in its canary implementation~\cite{vidurimpl}. LLMServingSim combines layerwise profiling with serving and interconnect simulation~\cite{llmservingsim}, including P--D handoff and device/CPU/CXL caching~\cite{llmservingsimdocs}. These approaches represent hardware and execution in detail. Our whole-iteration and effective-path calibration instead aligns service estimates with the target engine's execution boundaries; explicit KV dependencies compose them into cluster behavior.

\textit{Stateful workload modeling.} Frontier uses role-specific abstractions for disaggregated execution and stateful requests~\cite{frontier}. AgentServeSim models agent-program orchestration, including completion-driven successor release and retention across tool gaps~\cite{agentservesim}. Such dependencies address how workload demand evolves. Our open-loop coding replays hold recorded ordering and scaled interarrival intervals fixed, allowing cluster policies to be evaluated under identical offered demand.

Table~\ref{tab:related-scope} positions our focus on KV-constrained service composition. A reusable prefix affects latency through its location, destination capacity, and retrieval time. Representing these dependencies separately from policy selection and service prediction enables calibration reuse when cluster configurations change. Section~\ref{sec:evaluation} tests this capability across topologies and against an adapted LLMServingSim.

\begin{table}[t]
\caption{Selected modeling emphases of related simulators.}
\label{tab:related-scope}
\centering\footnotesize
\setlength{\tabcolsep}{3pt}
\begin{tabular}{@{}>{\raggedright\arraybackslash}p{0.30\columnwidth}>{\raggedright\arraybackslash}p{0.30\columnwidth}>{\raggedright\arraybackslash}p{0.32\columnwidth}@{}}
\toprule
Framework & Modeling emphasis & State or workload focus\\
\midrule
Vidur~\cite{vidur,vidurimpl} & Profiled operator costs & Serving; prefix caching in canary\\
LLMServingSim~\cite{llmservingsim,llmservingsimdocs} & Hardware--serving interaction & PD execution; tiered prefix caching\\
Frontier~\cite{frontier} & Role-specific execution & Stateful serving requests\\
AgentServeSim~\cite{agentservesim} & Program orchestration & Successor release and tool-gap retention\\
This work & KV-constrained service composition & Residency, reservations, and visibility\\
\bottomrule
\end{tabular}
\end{table}

\section{Simulator Architecture}
\label{sec:architecture}
\subsection{Overview}
\begin{figure*}[t]
\centering
\includegraphics[width=0.96\textwidth]{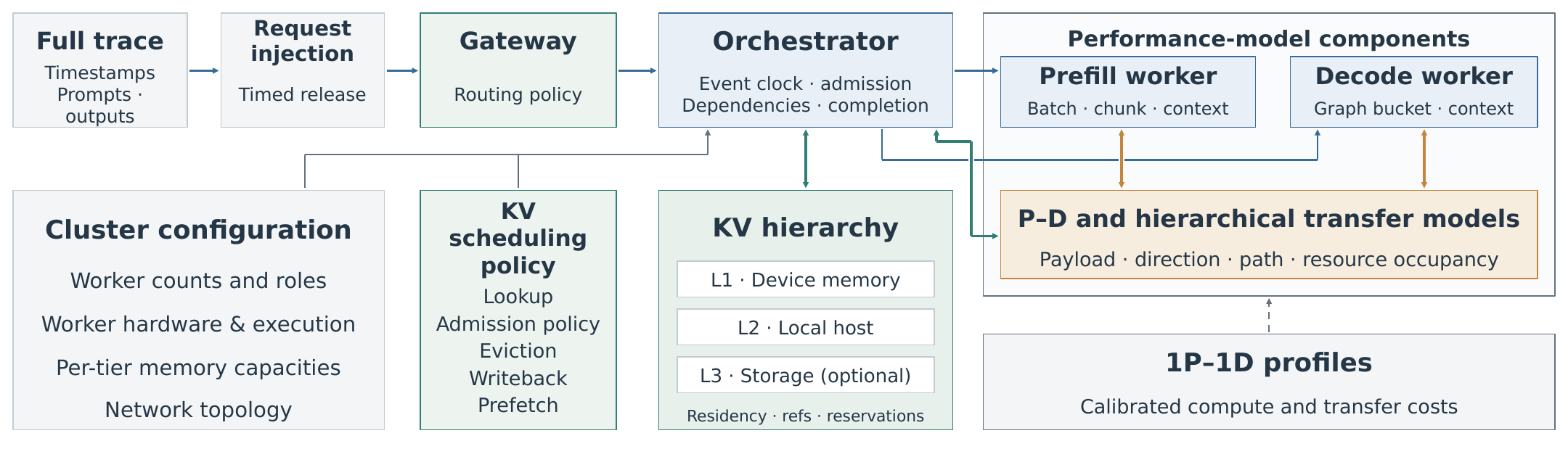}
\caption{Policies select work, service models predict duration, and shared KV state governs execution eligibility.}
\label{fig:overview}
\end{figure*}

To predict how serving policies affect cluster latency, the architecture separates decisions, service costs, and execution eligibility (Fig.~\ref{fig:overview}). Configuration and policy interfaces define the workload and its placement; hierarchical KV state determines data availability; local models supply operation durations. The orchestrator combines these inputs into completion events and traces of latency, queues, and reuse.

A host-prefix hit illustrates these dependencies: a request reserves device capacity and retrieves the prefix, computes its uncached suffix, and transfers KV to the decode worker. Contention at each stage determines the latency benefit of reuse.

\subsection{User Configuration and Policy Injection}
A cluster configuration defines the available workers, their execution settings, connectivity, and per-tier capacities. Workloads retain token identities for prefix matching, output lengths for generation, and session identifiers for affinity routing. Arrivals are independent of simulated completions, so alternative policies receive the same open-loop demand.

Gateway policies select workers, while KV policies select reuse and retention actions, including eviction, writeback, and prefetching. These decisions are proposals subject to the orchestrator's capacity and dependency checks. Separating policy selection from physical allocation preserves a common model of feasible execution across policy changes.

\subsection{Hierarchical KV Representation}
Logical KV content may have several physical replicas with different access costs. We represent each replica by its location, sharing scope, tier, allocation, and access path. Prefix matching identifies the reusable content, while replica state determines where it can be accessed and whether it can be reclaimed. P--D handoff creates availability at another worker through a transfer rather than an instantaneous change of ownership.

The representation also distinguishes allocated space from usable data. An in-progress copy reserves destination capacity but becomes available to computation only on completion. Each physical allocation is counted once even when several requests reference it. These rules prevent shared prefixes from being charged repeatedly and concurrent transfers from claiming the same free space.

\subsection{Performance-Model Components}
Service models predict the duration of an eligible operation from its workload and execution context. Prefill depends on executed tokens and existing context; decode depends on active sequences, context lengths, and CUDA-graph configuration; transfer depends on payload and access path. Section~\ref{sec:modeling} defines the corresponding predictors. The orchestrator accounts separately for admission and resource waiting.

Backends combine service estimates with resource queues and sharing rules to schedule completions. Profile changes affect service costs; backend changes may also affect contention. Both retain the orchestrator's completion interface, allowing profile reuse across compatible execution layouts and calibration domains.

\subsection{Discrete-Event Orchestration}
The orchestrator advances between arrivals, operation completions, and maintenance events. Each event updates state and releases any newly available resources before eligible work is dispatched. Queueing and overlap consequently emerge from dependencies and resource occupancy rather than a separately fitted queueing-delay model.

KV movement obeys three invariants: destination capacity is reserved before dispatch, source replicas remain protected while in use, and destination replicas become visible only on completion. Requests that cannot satisfy admission conditions remain pending; reservations are released if an operation cannot proceed. These invariants impose consistent capacity and availability constraints across policies and access paths.

\section{Data-Driven Modeling}
\label{sec:modeling}
\subsection{Calibration and Composition}
\begin{figure}[!ht]
\centering
\includegraphics[width=0.96\columnwidth]{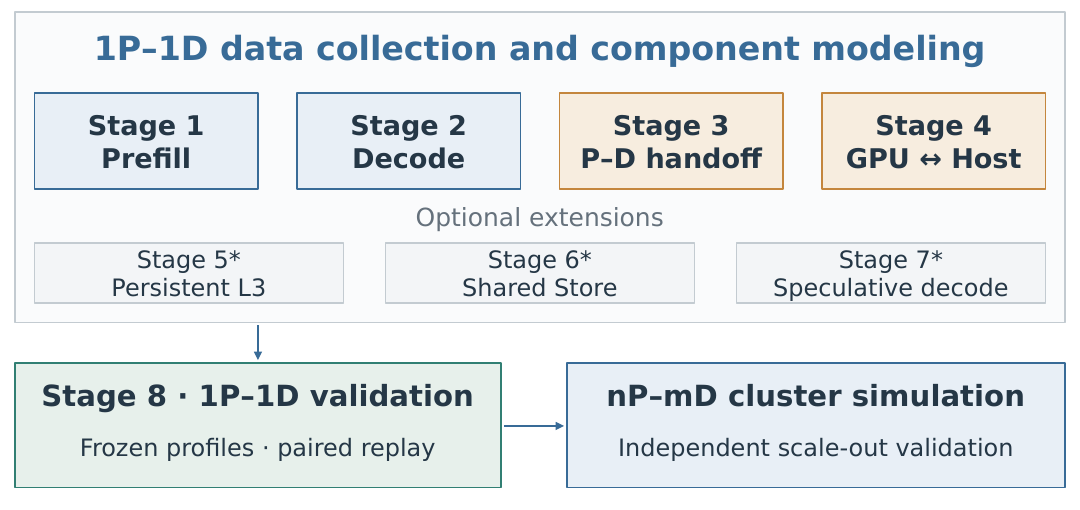}
\caption{Local compute and transfer profiles are assessed individually and composed for cluster-level validation.}
\label{fig:modeling-stages}
\end{figure}

The modeling workflow turns local measurements into reusable compute and transfer predictors (Fig.~\ref{fig:modeling-stages}). Calibration on a logical 1P--1D deployment fixes the model, hardware, and KV layout while varying operation sizes and batches. Profiles record timing boundaries, fitted parameters, and measured domains to support reuse across compatible workers.

We first compose the profiles in 1P--1D replay, where request timing depends on their interaction with queueing and KV state. Larger $n$P--$m$D configurations instantiate additional worker queues and capacities while retaining compatible profiles and modeling shared transfer resources. Changing cluster size therefore changes resource occupancy and event ordering without requiring an end-to-end latency fit.

Component tests assess service-time prediction, while cluster tests evaluate how those costs compose under admission, queueing, and multi-source contention. We use frozen profiles in both 1P--1D and 3P--1D replay, distinguishing interpolation within measured domains from extrapolation to other request shapes.

\subsection{Compute Modeling}
\textit{Prefill.} Prefix reuse reduces executed work but leaves cached tokens in the attention context. For request $i$, let $q_i$ be the tokens executed in the current step, $c_i$ the preceding device-resident context, and $n_i$ the original input length. Context $c_i$ includes reused prefixes and completed chunks; a lower-tier match contributes only after retrieval. For batch size $b$, we predict
\begin{equation}
 \widehat t_{\mathrm P}=f_{\mathrm P}\!\left(b,\{(q_i,c_i,n_i)\}_{i=1}^{b};\theta_{\mathrm P}\right).
\end{equation}
The coefficients $\theta_{\mathrm P}$ are fitted by regularized regression on features of executed tokens, attention work, and batch shape. Batch features include the maximum and dispersion of original input lengths, which remain fixed across steps. Distinguishing $q_i$ from $c_i$ accounts for the attention cost of a reused prefix without charging for its recomputation~\cite{sglang}. Calibration varies suffix lengths, prefix lengths, and batch composition to capture these effects.

\textit{Decode.} A standard decode iteration produces one token per active sequence. CUDA graphs reduce launch overhead by replaying captured operations~\cite{cudagraphs}, but the smallest configured bucket accommodating batch size $b$ also determines padding. We represent the resulting discrete changes using a GPU-time model for each bucket $g$:
\begin{equation}
 \widehat t_{\mathrm D}^{\mathrm{GPU}}=a_g+d_g\sum_{i=1}^{b}\ell_i,
\end{equation}
where $\ell_i$ is the context length of sequence $i$. The intercept $a_g$ captures fixed bucket costs; $d_g$ scales aggregate context length to approximate KV-dependent attention. Tests with homogeneous and mixed context lengths assess this aggregate approximation. A separate deployment correction maps GPU time to decode cadence; it is calibrated from previously collected light-load client time per output token (TPOT) and fixed before replay. Profiles correspond to the configured graph capture, hardware, and execution settings; the orchestrator adds queueing and KV retrieval to these compute costs.

\subsection{Transfer Modeling}
Transfer cost depends on the operation, direction, endpoints, and physical payload. The KV page allocation and worker execution layout determine how token counts map to bytes. For operation and path $k$, we use
\begin{equation}
 \widehat t_k(B)=\alpha_k+h_k(B),
\end{equation}
where $B$ is payload size, $\alpha_k$ is startup cost, and $h_k$ is a linear or piecewise-interpolated function. Each direction and path is calibrated separately. The backend combines this isolated service estimate with sharing rules to determine completion time; the orchestrator then makes the transferred KV available.

\textit{P--D handoff.} Calibration measures the interval from prefill completion to handoff completion on the prefill host. Because the interval includes software coordination, its fitted rate represents effective service rather than NIC bandwidth. In the evaluated configuration, each source processes one active chunk at a time in FIFO order. When $N$ chunks transfer concurrently to a decode worker, each receives $1/N$ of its isolated size-dependent rate. Multi-source cluster replay tests this sharing assumption beyond the pairwise calibration.

\textit{Cross-tier movement.} Separate host-to-device (H2D) and device-to-host (D2H) models capture prefix retrieval and writeback. The same interface can represent lower storage tiers through metadata queries and payload reads or writes: lookup cost depends on the metadata operation, while payload cost depends on bytes and access path. The evaluated profiles cover GPU--host movement. Recorded payload ranges distinguish interpolation from extrapolation.

\section{Evaluation}
\label{sec:evaluation}
We test calibration reuse through component and cluster accuracy, compare against LLMServingSim, and quantify cache fidelity and evaluation time.

\subsection{Experimental Setup}
\label{sec:setup}
We evaluate SGLang~\cite{sglang} serving Qwen3-8B and Qwen3-32B~\cite{qwen3} on RDMA-connected NVIDIA L20Z hosts. Comparing 1P--1D and 3P--1D varies prefill capacity with one decode worker. Each worker uses two GPUs with tensor parallelism (TP2) and 262144 device KV tokens; each P additionally provides 524352 host-memory tokens through HiCache~\cite{hicache}. Both topologies use round-robin routing, BF16 KV, 64-token pages, 4096-token prefill chunks, and a limit of 32 running requests per worker. D-side prefix and host caching, as well as L3, are disabled.

The workloads cover document reuse and session reuse (Table~\ref{tab:workloads}). QASPER~\cite{qasper} and SQuAD~\cite{squad} requests share prefixes within a paper or paragraph and generate 128 tokens. Hashed sampling retains complete document groups to preserve reuse opportunities. Coding traces~\cite{codingcaptures} (labeled Agentic in results) retain complete sessions whose lengths fall within the source dataset's interquartile range.
\begin{table}[!ht]
\caption{Workloads and mean request lengths. Groups are papers, paragraphs, and sessions, respectively.}
\label{tab:workloads}
\centering\fontsize{8}{9.3}\selectfont
\setlength{\tabcolsep}{2.5pt}
\begin{tabular}{@{}lrrrrl@{}}
\toprule
Trace & Groups & Reqs. & \multicolumn{2}{c}{Mean tokens} & Arrivals\\
 & & & Input & Output & \\
\midrule
QASPER & 129 & 506 & 5449.8 & 128.0 & Uniform\\
SQuAD & 71 & 400 & 226.0 & 128.0 & Uniform\\
Coding & 50 & 451 & 13325.4 & 650.4 & Trace-scaled\\
\bottomrule
\end{tabular}
\end{table}

A \emph{replay case} specifies a model, topology, dataset, and offered request rate. For each topology and QA dataset, rates are 1 and 2 requests/s for 8B, and 0.5 and 1 for 32B. Coding replays preserve request contents and output lengths while scaling recorded interarrival intervals to 0.25 and 0.5 requests/s for 8B, and 0.125 and 0.25 for 32B; 3P--1D additionally uses 1 and 0.5, respectively. The resulting 16 QA and ten coding cases each replay a complete dataset from empty caches. Table entries average per-case metrics equally within each row; QA rows combine both datasets. Cases share source requests rather than independent samples. QA datasets were prepared after model profiles and the simulator were frozen, whereas coding traces informed mechanism development.

\subsection{Component Accuracy}
Compute errors remain below 5\% and transfer errors below 10\% for both models (Table~\ref{tab:current-components}). Tests vary prefill suffix and context lengths, decode batch and context sizes, and transfer payloads. Training and test requests differ, although some writeback sizes recur; the 32B prefill result uses previously inspected development data.
\begin{table}[!ht]
\caption{Component service-time MAPE (\%): compute errors below 5\%, transfer errors below 10\%.}
\label{tab:current-components}
\centering\footnotesize
\begin{tabular}{@{}lrr@{}}
\toprule
Component & 8B & 32B\\
\midrule
Prefill GPU & 4.62 & 2.18\\
Decode GPU & 3.14 & 2.13\\
P--D effective completion & 3.79 & 9.58\\
Host $\rightarrow$ GPU & 6.97 & 4.15\\
GPU $\rightarrow$ Host & 0.19 & 0.17\\
\bottomrule
\end{tabular}
\end{table}

Cluster replays retain the prefill and transfer fits, with decode recalibrated for the enlarged KV pool. They test how these service estimates compose under admission, queueing, and overlapping transfers, including extrapolation to SQuAD prompts shorter than the compute calibration inputs.

\subsection{Cluster Accuracy}
\begin{figure}[!ht]
\centering
\includegraphics[width=\columnwidth]{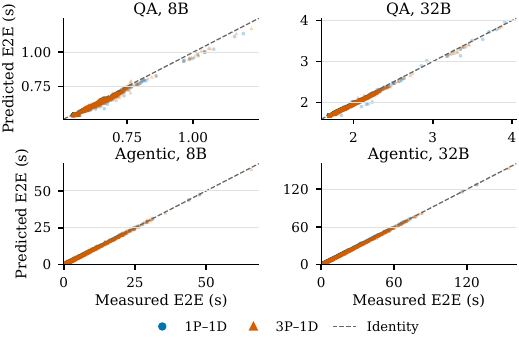}
\caption{E2E predictions across both topologies: mean MAPE is 1.34\% for QA and 6.36\% for coding sessions. Replay cases are equally weighted.}
\label{fig:e2e-validation}
\end{figure}

Mean E2E MAPE is 1.34\% for QA and 6.36\% for coding sessions, measuring latency from client request start to last token. Figure~\ref{fig:e2e-validation} includes every paired request in both topologies, supporting reuse of local profiles as prefill capacity increases. Coding sessions exercise bursty arrivals, variable generation lengths, and changing queues and batches.

We assess responsiveness and streaming separately in Table~\ref{tab:llmservingsim-comparison}. Time to first token (TTFT) spans client request start to first token; TPOT averages subsequent inter-token intervals. TTFT error uses mean absolute error (MAE), and TPOT error uses MAPE. QA errors remain similar across topologies; coding sessions have larger first-token errors alongside longer generations.

\begin{figure}[!ht]
\centering
\includegraphics[width=\columnwidth]{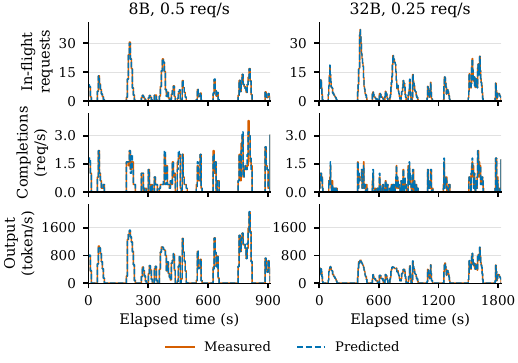}
\caption{Predicted and measured backlog and throughput follow similar temporal patterns in 5-s windows of 3P--1D coding replay (8B: 0.5 requests/s; 32B: 0.25).}
\label{fig:temporal-validation}
\end{figure}

Predictions also reproduce the main accumulation and draining patterns in complete 3P--1D coding replays (Fig.~\ref{fig:temporal-validation}). Aligned 5-s windows show outstanding requests, including waiting and execution, alongside completed requests and delivered tokens. These complementary views connect request latency to client-visible backlog and throughput.

\subsection{Comparison with LLMServingSim}
We compare against LLMServingSim~\cite{llmservingsim,llmservingsimdocs} on the same workloads with identical tokens, arrivals, routes, and KV capacities. The baseline uses measured L20Z kernel profiles, calibrated collective and host-transfer rates, and nominal RDMA bandwidth. We correct verified host-capacity accounting and cache-hit handling issues. Its native P--D path aborts on all ten coding cases; to complete these replays, we add reservation of decode capacity for the prompt and generation budget before prefill. The compute, batching, and caching models remain unchanged. This adapted baseline tests the transfer of a vLLM-based execution model to our SGLang deployments.

Our E2E and TTFT errors are consistently lower (Table~\ref{tab:llmservingsim-comparison}). Coding E2E MAPE averages 6.36\%, versus 15.21\% for the baseline, and coding TPOT errors are lower in every case. QA TPOT favors LLMServingSim for 8B and our simulator for 32B.
\begin{table}[!ht]
\caption{Mean errors by P:D topology (ours / adapted LLMServingSim): lower E2E and TTFT errors in every group.}
\label{tab:llmservingsim-comparison}
\centering\footnotesize
\setlength{\tabcolsep}{2.8pt}
\begin{tabular}{@{}lrrr@{}}
\toprule
Group & E2E MAPE (\%) & TTFT MAE (ms) & TPOT MAPE (\%)\\
\midrule
QA 8B 1:1 & 1.55 / 5.25 & 15.4 / 33.6 & 1.20 / 0.47\\
QA 8B 3:1 & 1.58 / 5.14 & 15.1 / 32.9 & 1.12 / 0.41\\
QA 32B 1:1 & 1.06 / 3.39 & 23.4 / 62.5 & 0.38 / 0.80\\
QA 32B 3:1 & 1.16 / 3.30 & 24.7 / 61.1 & 0.34 / 0.74\\
Agentic 8B 1:1 & 5.83 / 15.25 & 86.1 / 236.6 & 5.53 / 7.70\\
Agentic 8B 3:1 & 9.07 / 16.90 & 155.1 / 681.1 & 2.98 / 5.64\\
Agentic 32B 1:1 & 5.04 / 12.72 & 207.0 / 565.1 & 1.70 / 4.01\\
Agentic 32B 3:1 & 4.88 / 15.16 & 215.1 / 2118.9 & 1.69 / 4.37\\
\bottomrule
\end{tabular}
\end{table}

Our calibration captures whole iterations and effective transfer paths at the target engine's execution boundaries. LLMServingSim composes kernel and layer costs with its serving model. KV availability and resource queues translate these costs into request timing. The comparison evaluates these complete simulator configurations, including engine and admission differences, rather than isolating individual mechanisms.

\subsection{Prefix Cache Accuracy}
Mean token-hit-ratio error remains within 1.23 percentage points (pp) in every group in Table~\ref{tab:prefix-summary}. Token hit ratio measures reused tokens as a fraction of total input; exact match measures requests with identical predicted and measured prefix lengths. Together they assess aggregate reuse and request-level fidelity.
\begin{table}[!ht]
\caption{Prefix prediction captures reuse across P:D topologies; case means are equally weighted (pp: percentage points).}
\label{tab:prefix-summary}
\centering\fontsize{8}{9.3}\selectfont
\setlength{\tabcolsep}{2.5pt}
\begin{tabular}{@{}llcrrrr@{}}
\toprule
Data & Model & P:D & \multicolumn{2}{c}{Token hit (\%)} & Exact & Hit error\\
 & & & Measured & Predicted & match (\%) & (pp)\\
\midrule
QA & 8B & 1:1 & 63.78 & 63.93 & 99.80 & 0.15\\
QA & 8B & 3:1 & 40.44 & 40.44 & 100.00 & 0.00\\
QA & 32B & 1:1 & 63.78 & 63.93 & 99.80 & 0.15\\
QA & 32B & 3:1 & 40.44 & 40.44 & 100.00 & 0.00\\
Agentic & 8B & 1:1 & 80.61 & 81.44 & 90.58 & 0.84\\
Agentic & 8B & 3:1 & 58.64 & 58.96 & 97.49 & 0.32\\
Agentic & 32B & 1:1 & 79.39 & 80.61 & 81.49 & 1.23\\
Agentic & 32B & 3:1 & 57.98 & 58.56 & 94.97 & 0.58\\
\bottomrule
\end{tabular}
\end{table}

The simulator captures the lower reuse when round-robin routing distributes related requests across three P workers. Coding aggregate agreement is closer than request-level agreement, as the exact-match column shows. All eight QASPER cases exercise host retrieval, with mean token-volume error of 0.87\% and matching host attribution for 99.68\% of requests. Coding cases exercise writeback without retrieval, providing complementary coverage of host-cache behavior.

\subsection{Simulation Speed}
CPU simulation reduces replay turnaround by 15.33$\times$ on average across the evaluated configurations. Table~\ref{tab:speed-summary} compares client replay duration with CPU time for simulation and comparison on an Intel Xeon host using Python~3.11. Replay includes prescribed arrival gaps; startup and calibration are excluded. Rows average per-case ratios.
\begin{table}[!ht]
\caption{Replay/CPU turnaround: ratios are averaged per case; replay includes arrival gaps.}
\label{tab:speed-summary}
\centering\fontsize{8}{9.3}\selectfont
\setlength{\tabcolsep}{2.5pt}
\begin{tabular}{@{}llcrrr@{}}
\toprule
Data & Model & Runs & Replay (s) & CPU (s) & Mean ratio ($\times$)\\
\midrule
QA & 8B & 8 & 339.6 & 55.5 & 10.59\\
QA & 32B & 8 & 679.8 & 53.6 & 23.47\\
Agentic & 8B & 5 & 1180.6 & 143.1 & 8.36\\
Agentic & 32B & 5 & 2370.8 & 141.9 & 16.85\\
\bottomrule
\end{tabular}
\end{table}

CPU times remain similar across models despite longer 32B replays because simulation processes events without executing neural-network operations. This turnaround gain supports repeated policy comparisons with fixed profiles; it is not a GPU computation speedup.

\section{Applications}
\label{sec:applications}
The calibrated framework supports policy selection by predicting how reuse changes work, latency, and occupancy. We evaluate session affinity on real deployments and TTL retention in simulation, keeping workloads, hardware, and service models fixed.

\subsection{Session-Affinity Routing}
Session affinity preserves each session's initial worker assignment to promote local prefix reuse~\cite{sglangrouter}. We assign sessions cyclically and compare affinity with round-robin routing on the 3P--1D coding traces at 0.5 requests/s for 8B and 0.25 for 32B. Predictions are frozen before measurement, and each policy is measured once from empty caches.

The simulator captures the policy's main effect: lower prefill work with little change in E2E latency. Predicted uncached-input reductions are 50.3\% and 47.7\% for 8B and 32B, compared with measured reductions of 48.2\% and 46.1\%. Mean E2E changes are $-0.23\%$ and $+1.21\%$ in simulation, versus $+0.44\%$ and $+1.32\%$ in measurement. The small 8B changes have opposite signs. The overall pattern is consistent with unchanged generation demand at the shared decode worker, showing why reuse and latency should be evaluated jointly.

\begin{figure}[!ht]
\centering
\includegraphics[width=\columnwidth]{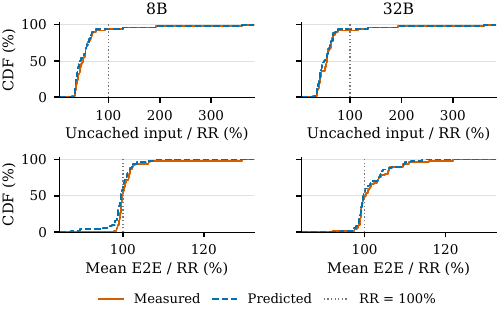}
\caption{Cumulative distributions over 50 sessions show lower aggregate uncached input with little E2E change under affinity. Ratios use round-robin (RR) = 100\%; 3P--1D rates are 0.5 requests/s (8B) and 0.25 (32B).}
\label{fig:session-affinity}
\end{figure}

Figure~\ref{fig:session-affinity} exposes variation across all 50 sessions, weighting sessions equally; aggregate latency means instead weight requests equally.

\subsection{KV Retention on Agent-Derived Traces}
TTL retention allows idle KV occupancy to be traded against later reuse. We augment least-recently-used (LRU) eviction with idle TTLs of 1, 10, 60, and 300\,s for P-side L1/L2 on the same coding traces. Periodic checks reclaim expired entries only when active references and transfers permit it. Capacity-driven eviction remains LRU, and nonexpiring controls reproduce the original predictions exactly.

\begin{figure}[!ht]
\centering
\includegraphics[width=\columnwidth]{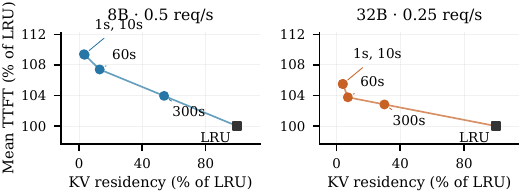}
\caption{Simulated idle TTLs reduce KV occupancy with a TTFT cost; both axes are normalized to LRU.}
\label{fig:ttl-case}
\end{figure}

A 60-s TTL reduces time-weighted KV occupancy by 86.9\% and 92.9\% for 8B and 32B, with mean TTFT increases of 7.4\% and 3.8\% relative to LRU (Fig.~\ref{fig:ttl-case}). All policies use a common request-injection window. Expiration releases idle slots but can remove prefixes before reuse, linking the trade-off to inter-request gaps and resource-release timing. Occupancy measures used KV slots; allocated GPU memory remains unchanged. The sweep lets designers compare retention settings against their tolerance for first-token delay.

Both studies reuse the same calibrated profiles, demonstrating policy substitution within a consistent model of KV availability, capacity, and contention.

\vfill
\section{Conclusion}
We presented a compositional simulator for evaluating LLM inference clusters with hierarchical KV caching. Explicit KV availability and resource queues enable local compute and transfer profiles to predict behavior across compatible workers. Across the evaluated deployments, E2E MAPE averages 1.34\% for QA and 6.36\% for coding; real replay takes 15.33$\times$ the CPU evaluation time on average. The framework supports early design-space exploration by quantifying how routing and retention policies jointly affect computational demand, request latency, and KV occupancy.

\clearpage
\bibliographystyle{IEEEtran}
\bibliography{references}

@inproceedings{vllm,
 author={Woosuk Kwon and Zhuohan Li and Siyuan Zhuang and Ying Sheng and Lianmin Zheng and Cody Hao Yu and Joseph E. Gonzalez and Hao Zhang and Ion Stoica},
 title={Efficient Memory Management for Large Language Model Serving with {PagedAttention}},
 booktitle={Proceedings of the ACM SIGOPS 29th Symposium on Operating Systems Principles},year={2023},
 url={https://arxiv.org/abs/2309.06180}}

@inproceedings{sglang,
 author={Lianmin Zheng and Liangsheng Yin and Zhiqiang Xie and Chuyue Sun and Jeff Huang and Cody Hao Yu and Shiyi Cao and Christos Kozyrakis and Ion Stoica and Joseph E. Gonzalez and Clark Barrett and Ying Sheng},
 title={{SGLang}: Efficient Execution of Structured Language Model Programs},
 booktitle={Advances in Neural Information Processing Systems},volume={37},year={2024},
 url={https://proceedings.neurips.cc/paper_files/paper/2024/hash/724be4472168f31ba1c9ac630f15dec8-Abstract-Conference.html}}

@inproceedings{distserve,
 author={Yinmin Zhong and Shengyu Liu and Junda Chen and Jianbo Hu and Yibo Zhu and Xuanzhe Liu and Xin Jin and Hao Zhang},
 title={{DistServe}: Disaggregating Prefill and Decoding for Goodput-optimized Large Language Model Serving},booktitle={18th USENIX Symposium on Operating Systems Design and Implementation},year={2024},url={https://arxiv.org/abs/2401.09670}}

@inproceedings{mooncake,
 author={Ruoyu Qin and Zheming Li and Weiran He and Jialei Cui and Feng Ren and Mingxing Zhang and Yongwei Wu and Weimin Zheng and Xinran Xu},
 title={{Mooncake}: Trading More Storage for Less Computation---A {KVCache}-centric Architecture for Serving {LLM} Chatbot},
 booktitle={23rd USENIX Conference on File and Storage Technologies},year={2025},pages={155--170},
 url={https://www.usenix.org/conference/fast25/presentation/qin}}

@inproceedings{vidur,
 author={Amey Agrawal and Nitin Kedia and Jayashree Mohan and Ashish Panwar and Nipun Kwatra and Bhargav S. Gulavani and Ramachandran Ramjee and Alexey Tumanov},
 title={{Vidur}: A Large-Scale Simulation Framework for {LLM} Inference},booktitle={Proceedings of Machine Learning and Systems},volume={6},year={2024},url={https://proceedings.mlsys.org/paper_files/paper/2024/hash/b74a8de47d2b3c928360e0a011f48351-Abstract-Conference.html}}

@inproceedings{llmservingsim,
 author={Jaehong Cho and Hyunmin Choi and Guseul Heo and Jongse Park},
 title={{LLMServingSim 2.0}: A Unified Simulator for Heterogeneous and Disaggregated {LLM} Serving Infrastructure},booktitle={IEEE International Symposium on Performance Analysis of Systems and Software},year={2026},pages={1--14},doi={10.1109/ISPASS69572.2026.00012}}

@article{frontier,
 author={Yicheng Feng and Xin Tan and Yangtao Deng and Yimin Jiang and Yibo Zhu and Hong Xu},
 title={{Frontier}: Towards Comprehensive and Accurate {LLM} Inference Simulation},journal={arXiv preprint arXiv:2605.21312},year={2026},note={Version 2},url={https://arxiv.org/abs/2605.21312v2}}

@article{agentservesim,
 author={Rakibul Hasan Rajib and Mengxin Zheng and Qian Lou},
 title={{AgentServeSim}: Serving-System Simulation and Policy Search for {LLM} Agent Programs},journal={arXiv preprint arXiv:2606.09613},year={2026},note={Version 3},url={https://arxiv.org/abs/2606.09613v3}}

@article{qwen3,
 author={An Yang and Anfeng Li and Baosong Yang and Beichen Zhang and Binyuan Hui and Bo Zheng and others},
 title={{Qwen3} Technical Report},journal={arXiv preprint arXiv:2505.09388},year={2025},url={https://arxiv.org/abs/2505.09388}}

@inproceedings{qasper,
 author={Pradeep Dasigi and Kyle Lo and Iz Beltagy and Arman Cohan and Noah A. Smith and Matt Gardner},
 title={A Dataset of Information-Seeking Questions and Answers Anchored in Research Papers},
 booktitle={Proceedings of NAACL-HLT},year={2021},pages={4599--4610},
 doi={10.18653/v1/2021.naacl-main.365},url={https://aclanthology.org/2021.naacl-main.365/}}

@inproceedings{squad,
 author={Pranav Rajpurkar and Jian Zhang and Konstantin Lopyrev and Percy Liang},
 title={{SQuAD}: 100,000+ Questions for Machine Comprehension of Text},
 booktitle={Proceedings of EMNLP},year={2016},pages={2383--2392},
 doi={10.18653/v1/D16-1264},url={https://aclanthology.org/D16-1264/}}

@misc{codingcaptures,
 author={{dacorvo}},title={Transformers Coding Session Captures},year={2026},
 howpublished={Hugging Face dataset},note={Revision ebf80ae3016c39616ccb08066198102eb8b16ed6},
 url={https://huggingface.co/datasets/dacorvo/transformers-coding-session-captures}}

@misc{vidurimpl,
 author={{Microsoft}},title={{Vidur}: Official Implementation},
 howpublished={GitHub repository},note={Accessed September 20, 2026; README documents canary prefix caching},
 url={https://github.com/microsoft/vidur}}

@misc{llmservingsimdocs,
 author={{CASYS, KAIST}},title={{LLMServingSim}: Official Documentation},
 howpublished={Official documentation},note={Accessed September 20, 2026},
 url={https://llmservingsim.ai/docs/getting-started/overview}}

@misc{sglangrouter,
 author={{SGLang Project}},title={{SGLang Model Gateway}: Routing Policy Configuration},
 howpublished={Official source code},note={Accessed September 20, 2026},
 url={https://github.com/sgl-project/sglang/blob/main/sgl-model-gateway/src/config/types.rs}}

@inproceedings{orca,
 author={Gyeong-In Yu and Joo Seong Jeong and Geon-Woo Kim and Soojeong Kim and Byung-Gon Chun},
 title={{Orca}: A Distributed Serving System for Transformer-Based Generative Models},
 booktitle={16th USENIX Symposium on Operating Systems Design and Implementation},
 year={2022},pages={521--538},url={https://www.usenix.org/conference/osdi22/presentation/yu}}

@inproceedings{sarathi,
 author={Amey Agrawal and Nitin Kedia and Ashish Panwar and Jayashree Mohan and Nipun Kwatra and Bhargav Gulavani and Alexey Tumanov and Ramachandran Ramjee},
 title={Taming Throughput-Latency Tradeoff in {LLM} Inference with {Sarathi-Serve}},
 booktitle={18th USENIX Symposium on Operating Systems Design and Implementation},
 year={2024},pages={117--134},url={https://www.usenix.org/conference/osdi24/presentation/agrawal}}

@misc{cudagraphs,
 author={{NVIDIA}},title={{CUDA} Programming Guide: {CUDA} Graphs},
 howpublished={Official documentation},note={Accessed September 20, 2026},
 url={https://docs.nvidia.com/cuda/cuda-programming-guide/04-special-topics/cuda-graphs.html}}

@inproceedings{flexgen,
 author={Ying Sheng and Lianmin Zheng and Binhang Yuan and Zhuohan Li and Max Ryabinin and Beidi Chen and Percy Liang and Christopher Re and Ion Stoica and Ce Zhang},
 title={{FlexGen}: High-Throughput Generative Inference of Large Language Models with a Single {GPU}},
 booktitle={Proceedings of the 40th International Conference on Machine Learning},
 volume={202},pages={31094--31116},year={2023},publisher={PMLR},
 url={https://proceedings.mlr.press/v202/sheng23a.html}}

@misc{hicache,
 author={{SGLang Project}},title={{HiCache} System Design and Optimization},
 howpublished={Official documentation},note={Accessed September 20, 2026},
 url={https://docs.sglang.io/docs/advanced_features/hicache_design}}
\end{document}